 \documentclass[journal]{IEEEtran}

\usepackage{amsmath,amsfonts}
\usepackage{algorithm}
\usepackage{algorithmic}
\usepackage{array}
\usepackage{bm}
\usepackage{amsmath,amssymb}
\usepackage[compress]{cite}
\usepackage[caption=false,font=normalsize,labelfont=sf,textfont=sf]{subfig}
\usepackage{textcomp}
\usepackage{stfloats}
\usepackage{url}
\usepackage{verbatim}
\usepackage{changebar}
\usepackage{graphicx}
\usepackage{rotating}
\usepackage{color}
\usepackage{hyperref} 

\def\BibTeX{{\rm B\kern-.05em{\sc i\kern-.025em b}\kern-.08em
    T\kern-.1667em\lower.7ex\hbox{E}\kern-.125emX}}
\usepackage{balance}

\begin{document}

\title{\huge FlexEdge: Digital Twin-Enabled Task Offloading for UAV-Aided Vehicular Edge Computing}
\author{Bin Li, Wancheng Xie, Yinghui Ye, Lei Liu, and Zesong Fei,~\IEEEmembership{Senior Member,~IEEE}
	
\thanks{B. Li and W. Xie are with the School of Computer Science and the Jiangsu Collaborative Innovation Center of Atmospheric Environment and Equipment Technology (CICAEET), Nanjing University of Information Science and Technology, Nanjing, 210044 (e-mail: bin.li@nuist.edu.cn; zuoyeyiwancheng@gmail.com).}
\thanks{Y. Ye is with the Shaanxi Key Laboratory of Information Communication Network and Security, Xi’an University of Posts \& Telecommunications, Xi'an 710121, China (e-mail: connectyyh@126.com).}
\thanks{L. Liu is with the Guangzhou Institute of Technology, Xidian University, Guangzhou 510555, China (e-mail: tianjiaoliulei@163.com).}
\thanks{Z. Fei is with the School of Information and Electronics, Beijing Institute of Technology, Beijing 100081, China (e-mail:feizesong@bit.edu.cn).}

}
\maketitle

\begin{abstract}
	
Integrating unmanned aerial vehicles (UAVs) into vehicular networks have shown high potentials in affording intensive computing tasks. 
In this paper, we study the digital twin driven vehicular edge computing networks for adaptively computing resource management where an unmanned aerial vehicle (UAV) named \textit{FlexEdge} acts as a flying server. In particular, we first formulate an energy consumption minimization problem by jointly optimizing UAV trajectory and computation resource under the practical constraints.
To address such a challenging problem, we then build the computation offloading process as a Markov decision process and propose a deep reinforcement learning-based proximal policy optimization algorithm to dynamically learn the computation offloading strategy and trajectory design policy. Numerical results indicate that our proposed algorithm can achieve quick convergence rate and significantly reduce the system energy consumption.

\end{abstract}


\section{Introduction}
Internet of Vehicles are expected to play a critical role in future digital cities such as smart driving and intelligent transportation systems \cite{2022ZhangTII_Ada}. 
Considering the limited computing resource on the vehicles, vehicular edge computing (VEC) is recognized as a promising solution to enable vehicular real-time services via offloading computation-intensive tasks to the network edge  \cite{Dujianbo2020,2022YangTVT_Intelligent}.  
Generally, road side units (RSUs) serve as the edge nodes to provide computation and communication resources for the vehicles running on the road. 
However, the highly dynamic topology of vehicular networks may make the effective interaction time duration of both vehicle-to-vehicle and vehicle-to-RSU extremely short. Furthermore, the locations of RSUs are usually fixed, and the deployment of MEC servers requires a certain amount of space and cost.

Recently,  unmanned aerial vehicle (UAV)-assisted VEC has drawn extensive attention due to the provided ubiquitous connectivity and three-dimensional networking coverage for realizing the task offloading \cite{2021JiangIN_AI}. Compared with the traditional VEC where the computing facilities are only available at RSUs, the UAV-assisted VEC can provide flexible services according to the actual road conditions and mission requirements \cite{2021HuICSM_U,2021PengJSAC_M}. 

In practical VEC, how to design an appropriate mechanism to optimize the offloading decisions is a challenge \cite{DuTNSE2022}. As a potential cure, digital twin (DT) has recently proposed to build virtual network space and provide virtual images of corresponding physical entities \cite{KhanSurvey}. Based on this architecture, DT can replace the vehicles and edge servers to make offloading decisions in the virtual space in advance, while the computing and communication resources between vehicles and edge servers in physical space can be provided quickly and accurately according to the request of the vehicles \cite{2022LiIAP_unmanned}. This is of paramount significance to capture the time-varying resource supply and demand in the development of VEC \cite{HuynhTCOM,2022ChenTVT_E}. 

Recently, many research efforts have mainly focused on DT-aided service architecture. In particular, Zhang \emph{et al.} \cite{2022ZhangTII_Ada} integrated DT with multiagent learning to optimize edge resource scheduling in VEC networks. Dai \emph{et al.} \cite{2021DaiTII_Deep} introduced DT to model the stochastic task arrival and leveraged asynchronous actor-critic to minimize the energy consumption. With the support of DT technology, the intelligent offloading with edge selection was studied in \cite{2022TanLWC_Digital}, while integrate computing, communication, and storage was considered in \cite{2022VanLWC_Edge} to minimize the latency performance. Yuan \emph{et al.} \cite{YuanTITS2022} proposed a dynamic DT of the VEC network to reflect the network characteristics in real-time. To provide the seamless coverage and high-quality services, Li \emph{et al.} \cite{2022LiTVT_Digital} exploited DT to support UAV-enabled MEC systems where deep Q-network is proposed. 

Although the aformentioned excellent studies laid an initial foundation on DT-aided MEC, the application of DT in UAV-assisted VEC networks to help vehicles making the offloading decisions has not been considered.
We in this paper propose a new DT architecture to facilitate the computation offloading in UAV-aided VEC network. Our specific contributions
	are:
\begin{itemize}
	\item We introduce DT to VEC networks for achieving real-time computing, where UAV has two roles to play: aerial edge server and mobile relay. Specifically, the vehicles can offload part of the computing tasks to the UAV or to the RSU via UAV relay link for edge processing. The deviation between the estimated computing frequency and the real value of devices is carefully considered. 

	\item The formulated energy consumption minimization optimization problem is a hybrid discrete-continuous action space problem and the offloading decision and UAV trajectory are also closely coupled with each other. We formulate the vehicles and UAV status update problem as a Markov decision process (MDP) and leverage the online proximal policy optimization (PPO) algorithm to learn environment dynamics and computing demands via DT in order to enable real-time offloading decisions and UAV trajectory policy.
	 
\end{itemize}


\section{System Model and Problem Statement} 

\begin{figure}[t]
	\centering
	\includegraphics[width=0.40\textwidth]{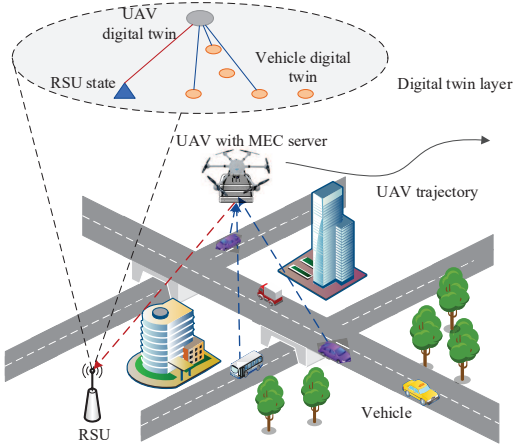}
	\caption{Digital twin model for UAV-assisted VEC.}
	\label{fig:sys-model}
\end{figure}

We consider a UAV-assisted vehicular network containing $K$ vehicles, an RSU and a UAV, as shown in Fig.~\ref{fig:sys-model}. To better express the system state, we introduce a time period $T$, which is spanided into $N$ time slots, where the length of each time slot is denoted as $\delta_t=T/N$. We define the set of vehicles and the set of time slots as $\mathcal{K}=\{1,\ldots,K\}$ and $\mathcal{N}=\{1,\ldots,N\}$, respectively. Since the vehicles typically have limited computing resource, they need to offload the time-sensitive tasks to the RSU equipped with VEC server for further processing. 
However, the communication signal between the vehicles and RSU may be blocked by the obstacles (e.g., high building). In addition, RSU may far away from the vehicles, which makes it hostile for vehicles to directly communicate with RSU via poor signals or even interruption links. 
Combined with the RSU, UAV works as the temporary edge server or moving relay to provide timely communication and computation services for the vehicles. In this case, the vehicles can offload the portion of tasks to UAV server or further to RSU server via UAV relay link. To timely evaluate the status of the network, the DT layer is maintained at the central controller to manage UAV and RSU resources. For ease of exposition, the locations of vehicles, UAV, and RSU at time slot $n$ are given by $\bm{w}_k[n]=[x_k[n],y_k[n],0]^{\text{T}}$, $\bm{q}[n]=[x_u[n],y_u[n],H]^{\text{T}}$, and $\bm{w}_r[n]=[x_r[n],y_r[n],0]^{\text{T}}$, respectively, where $H$ is the flying altitude of UAV.

We employ the orthogonal frequency division multiple access protocol to avoid the interference between vehicles. Hence, the uplink transmission rate from vehicle $k$ is given by
\begin{equation}
R_k[n]=\frac{B}{K}\log_2\left(1+ K\frac{p_kh_k[n]}{B N_0}\right),\label{eq:R_k}
\end{equation}
where $B$ is the total available bandwidth of the network, and $h_k[n]$ denotes the channel gain between vehicle $k$ and UAV, which is calculated by $h_k[n]=\beta_0/(\Vert \bm{q}[n]-\bm{w}_k[n] \Vert)^2$.

\subsection{Digital Twin Model}
The central controller periodically collects vehicles' and UAV' data to record the DT models. Several open-source platforms, including Eclipse Ditto \cite{KhanSurvey}, Model Conductor-eXtended Framework, Mago3D, DeepSense 6G \cite{Deepsense} and DeepVerse 6G \cite{Deepverse} have been designed for creating the DT-based systems. This can enable us to execute the digital system and physical system.

In this paper, DT is utilized not only to model the computing resources of vehicles and UAV server, but also to assist the model training and parameter synchronization of learning-based methods. 
For the $k$‐th vehicle, the virtual twin need to record its task information and location, which can be given by
\begin{equation}
	DT_k[n]=\Big\{V_k[n],\tilde{f}_k[n],\bm{w}_k[n]\Big\},
\end{equation}
where $V_k[n]=\{D_k[n],C_k[n],t_k[n]\}$ means the computation task with a latency requirement of $t_k[n]$, an input size of $D_k[n]$ bits, and an average number of central process unit (CPU) cycles to processing one bit data $C_k[n]$. Although DT model represents the operating state of the real network as accurately as possible, there are still mapping errors due to the limitations of the DT modeling method and the acquisition of modeling data. Hence, we denote $\tilde{f}_k[n]$ as the estimated CPU frequency for physical vehicle $k$ at time slot $n$.

For UAV, the DT needs to reflect its scheduling of service, involving the allocation of resource and location information. Thus, the virtual twin of UAV is expressed as
\begin{equation}
	DT_u[n]=\Big\{\tilde{f}_k^u[n],\bm{q}[n],\bm{a}[n],\bm{v}[n]\Big\},
\end{equation}
where $\tilde{f}_k^u[n]$ denotes the estimated CPU frequency for allocating to vehicle $k$ by UAV edge server, $\bm{a}[n]$ indicates the acceleration of UAV, and $\bm{v}[n]$ denotes the velocity of UAV at time slot $n$. The DT of UAV edge server monitors the current status of the physical edge servers and vehicles for subsequent real-time decision making.

\subsection{Computation Model}
At time slot $n$, each vehicle generates a task $V_k[n]$. We consider that the tasks can be divided into two parts and concurrently executed. Denoting $\alpha_k[n]$ as the task partition factor, which means that $\alpha_k[n]D_k[n]$ bits of task is computed at UAV or RSU, and $(1-\alpha_k[n])D_k[n]$ is computed locally.

1) \textit{Local computing}: The estimated local computing time is calculated as
\begin{equation}
	\tilde{T}_k^l[n]=(1-\alpha_k[n])D_k[n]/\tilde{f}_k[n]
\end{equation}
 
According to \cite{2022TanLWC_Digital} and \cite{YuanTITS2022}, the local computing time gap between real value and DT estimation can be given by
\begin{equation} 
	\Delta T_k^l[n]=\frac{-(1-\alpha_k[n])D_k[n]C_k[n]{\hat{f}_k[n]}}{\tilde{f}_k[n](\tilde{f}_k[n]+{\hat{f}_k[n]})},
\end{equation}
where $\hat{f}_k[n]$ denotes the estimated deviation of actual frequency $f_k[n]=\tilde{f}_k[n]+\hat{f}_k[n]$. Then, the actual local computing time is given by 
\begin{equation}
	T_k^l[n]=\tilde{T}_k^l[n]+\Delta T_k^l[n]
\end{equation}

2) \textit{Edge Computing}: In terms of edge computing, the procedure can be divided to three parts. First, the vehicles transmit the tasks to UAV. Then, the UAV receives and processes the tasks. Additionally, if the tasks cannot be completed, the UAV will relay some proportion of tasks to RSU for completing the tasks. We assume that the computing results are with small size, and thus the downloading time is negligible. 
Denoting $T_k^o[n]$ as the offloading time of vehicle $k$ at time slot $n$, which is calculated by $T_k^o[n]=\alpha_k[n]D_k[n]/R_k[n]$. $R_k[n]$ is the transmission rate according to the Shannon formula.

To this end, the computing energy of UAV is given by
\begin{align}
	\nonumber E_k^u[n]=&\kappa f_k^u[n]^2 \min\Big\{f_k^u[n] (t_k[n]-T_k^o[n]),\\
	&\alpha_k[n] D_k[n] C_k[n]\Big\}
\end{align}

It is worth noting that if the task of vehicle $k$ cannot be completed at UAV, the rest part  will be relayed to RSU concurrently. Admittedly, remote offloading helps to reduce the UAV’s energy consumption as some tasks are executed by the RSU. Note that the relay energy consumption is relatively negligible compared to the computing and flying energy of UAV. 
The relay time $T_k^r[n]$ can be calculated by the size of remain tasks $D_k^r[n]=\alpha_k[n]D_k[n]-f_k^u[n]T_k^o[n]/C_k[n]$ and transmission rate. Thus, we have 
\begin{equation}
	R^r[n]=\frac{B}{K}\log_2\left(1+K\frac{p_u h_u[n]}{B N_0}\right),
\end{equation}
and
\begin{equation}
	T_k^r[n]= D_k^r[n]/R^r[n].
\end{equation}

When a task is offloaded to the UAV edge server,
the computing time gap between real value $T_k^u[n]=\tilde{T}_k^u[n]+\Delta T_k^u[n]$ and estimated DT value $\tilde{T}_k^u[n]=\alpha_k[n]D_k[n]C_k[n]/\tilde{f}_k^u[n]$ is
\begin{equation} \label{eq:delta_Tu}
	\Delta T_k^u[n]=\frac{-\alpha_k[n]D_k[n]C_k[n]{\hat{f}_k^u[n]}}{\tilde{f}_k^u[n](\tilde{f}_k^u[n]+{\hat{f}_k^u[n]})},
\end{equation}
where $\hat{f}_k^u[n]$ is the estimated deviation of edge server actual frequency $f_k^u[n]=\tilde{f}_k^u[n]+\hat{f}_k^u[n]$.

Denoting the $\tilde{f}_k^{\rm rc}[n]$ and $\hat{f}_k^{\rm rc}[n]$ as the estimated CPU frequency allocated to vehicle $k$ and the estimated deviation of actual frequency $f_k^{\rm rc}[n]=\tilde{f}_k^{\rm rc}[n]+\hat{f}_k^{\rm rc}[n]$, the computing energy of RSU is calculated by 
\begin{align}
	 E_k^{\rm rc}[n]=&\kappa f_k^{\rm rc}[n]^2 \min\Big\{f_k^{\rm rc}[n] (t_k[n]-T_k^r[n]), D_k^r[n] C_k[n]\Big\}
\end{align}

Therefore, the estimated RSU computing time $\tilde{T}_k^{\rm rc}[n]$ and its estimated deviation $\Delta{T}_k^{\rm rc}[n]$ for real computing time $T_k^{\rm rc}[n]$ can be similarly calculated as \eqref{eq:delta_Tu} with the relayed task size $D_k^r[n]$.
Then, the actual latency of edge computing can be written as
\begin{equation}
	T_k^e[n]=T_k^o[n]+\max\Big\{T_k^r[n]+T_k^{\rm rc}[n],T_k^u[n]\Big\}
\end{equation}

\subsection{UAV Flying Model}
In each time slot $n$, the UAV flies obeying the constraints of speed and acceleration, which can be formulated as
\begin{equation}
	\Vert\bm{v}[n]\Vert\leq v_{\max},\forall n \in \mathcal{N}\label{c:v}.
\end{equation}
\begin{equation}
	\Vert\bm{a}[n]\Vert\leq a_{\max},\forall n\in\mathcal{N}\label{c:a},
\end{equation}
\begin{equation}
	\bm{q}[n+1]=\bm{q}[n]+\bm{v}[n]\delta_t+\frac{1}{2}\bm{a}[n]\delta_t^2,\forall n \in \mathcal{N}.
	\label{c:q}
\end{equation}
Then, the propulsion energy of UAV can be expressed as follows:
\begin{align}
	\nonumber E^f[n]=&\frac{1}{2}d_0\rho s A \Vert\bm{v}[n]\Vert^3+P_0\left(1+\frac{3\Vert \bm{v}[n]\Vert^3}{U_{\text{tip}}^2}\right)\\&+P_i\left(\sqrt{1+\frac{\Vert\bm{v}[n]\Vert^4}{4v_0^4}}-\frac{\Vert\bm{v}[n]\Vert^2}{2v_0^2}\right),\label{energy_fly}
\end{align}
where $P_i$ and $P_0$ are the induced power in hovering status and the blade power of UAV, $v_0$ is the mean rotor velocity, $U_{{\text{tip}}}$ denotes the tip speed of the blade, $d_0$ is the fuselage drag ratio, $s$ is the rotor solidity, $\rho$ denotes the air density, and $A$ is the rotor disc area.

\subsection{Problem Statement}
We aim to minimize the energy consumption of UAV and RSU, the optimization problem can be expressed as 
\begin{subequations}
	\begin{align}
		\min\limits_{\bm{\alpha},\bm{f},\bm{q}} ~~&\sum\limits_{n=1}^N \left(\sum\limits_{k=1}^K (E_k^u[n]+E_k^{\rm rc}[n])+E^f[n]\right)
		\hphantom{\min\limits_{\bm{\alpha},\bm{f}^u,\bm{q}}} \label{c:obj}\\
		\text{s.t.}\quad&\eqref{c:v},\eqref{c:a},\eqref{c:q},\label{c:move}\\
		&\max\{T_k^l[n],T_k^e[n]\} \leq t_k[n],\forall k \in \mathcal{K},\forall n \in \mathcal{N},\label{c:T}\\
		&0\leq\alpha_k[n]\leq 1,\forall k \in \mathcal{K},\forall n \in \mathcal{N},\label{c:alpha}\\
		&\sum\limits_{k=1}^K  \tilde{f}_k^u[n] \leq f_{\max}^u,\forall k \in \mathcal{K},\forall n \in \mathcal{N},\label{c:f_max}\\
		&\tilde{f}_k^u[n] \geq 0,\forall k \in \mathcal{K},\forall n \in \mathcal{N}.\label{c:f_0}
	\end{align}
	\label{c:opt}
\end{subequations}
where the optimization variables $\bm{\alpha}=\{\alpha_k[n]\}$, $\bm{f}=\{\tilde{f}_k[n],\tilde{f}_k^u[n],\tilde{f}_k^{\rm rc}[n]\}$, $\bm{q}=\{\bm{q}[n]\}$, $\forall k \in \mathcal{K},\forall n \in \mathcal{N}$. $f_{\max}^u$ is the maximum available CPU frequency of UAV in DT model. Constraint \eqref{c:move} represents the movement constraints of UAV. Constraint \eqref{c:T} ensures that the task execution time cannot exceed the maximum tolerable latency. Constraint \eqref{c:alpha} specifies the range of offloading proportion. Constraints \eqref{c:f_max} and \eqref{c:f_0} refer to the estimated computation resources for allocating to vehicle $k$ in digital space.

\section{Proposed DRL Approach}
In this section, we propose a PPO-based algorithm framework to address problem \eqref{c:opt} with dynamic communication states and highly-coupled variables. 

\subsection{DRL Components}
According to the general interaction model between DRL agent and network environment, the elements of MDP include state, action, and reward, which are defined as follows.
\begin{itemize}
	\item \textit{State}: In each time slot $n$, the DRL agent observes the state of the environment, which is presented by a four-tuple as $s_n=\{\bm{w}_k[n],\bm{q}[n],D_k[n],C_k[n]\},\forall k \in \mathcal{K}$. 
	\item \textit{Action}: After observing the state $s_n$, the agent executes an action $a_n=\{\alpha_k[n],f_k^u[n],f_k^{\rm rc}[n],\bm{q}[n]\},\forall k \in \mathcal{K}$, thus scheduling the resource and making offloading decisions for the UAV and the vehicles.
	\item \textit{Reward}: The agent executes the action based on the observed state and obtains an immediate reward $r_n$ from the environment. To reflect the optimization objective of \eqref{c:obj} in a long run, we design the form of the reward function similar to the system energy consumption.
	The reward consists of the system energy consumption and the penalty for violating the delay constraint, which is given by $r_n= \sum\limits_{k=1}^K (E_k^u[n]+E_k^{\rm rc}[n])+E^f[n] + P^l_n$, where $P^l_n=
	\frac{\mu}{K}
	\sum\limits_{k=1}^K\left(\max\Big\{T_k^l[n]-t_k[n],T_k^e[n]-t_k[n],0\Big\} \right)$ is a linear penalty function related to the violation degree $o_n$ of the latency constraint that is not satisfied, and $\mu$ is a coefficient of the penalty term.
\end{itemize}

\subsection{Learning Algorithm Design}
\begin{figure}[t]
	\centering
	\includegraphics[width=0.4\textwidth]{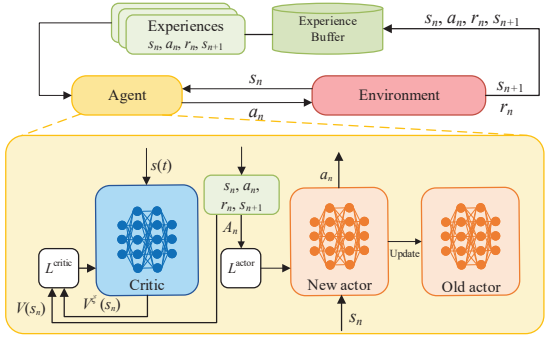}
	\caption{The framework of PPO algorithm.}
	\label{fig:PPO}
\end{figure}

In this subsection, we introduce the details of our proposed PPO algorithm. Here, the information involving MDP elements is uploaded and gathered in DT layer.
It can be readily observed that the state, action, and reward are continuous variables. Therefore, we leverage the PPO algorithm to approximate the optimal policy rather than discretizing the action and state spaces. The framework of PPO-based DRL training framework is displayed in Fig.~\ref{fig:PPO}.  Specifically, the PPO is based on actor-critic framework, where the actor network is used as policy to generate action $a_n$, and the critic network is used to evaluate the value of state $V(s_n)$ to adjust the current policy.

Different from the trust region policy optimization, PPO introduces a clipping factor in its objective function to restrict the update rate. Moreover, the objective of actor is optimized by the advantage function using generalized advantage estimator (GAE) that can effectively reduce the variance of gradient estimation, thus reducing the samples needed for training, there holds
\begin{equation}
	A_n=\sum\limits_{l=0}^\infty(\gamma\lambda)^l\Big(r_n+\gamma V(s_{n+1})-V(s_n)\Big),
\end{equation}
where $\gamma$ is the discount factor and $\lambda$ is GAE factor realizing a bias-variance tradeoff. In this paper, we develop a clip based PPO algorithm to train the actor-critic network. The probability ratio between the new policy and old policy is defined as $\Upsilon_\theta=\frac{\pi_{\theta}(a_n|s_n)}{\pi_{\theta^{'}}(a_n|s_n)}$, where $\theta$ and $\theta^{'}$ are the policy parameters concerning actor network and old actor network.
Accordingly, the loss function of the actor network is expressed as
\begin{equation}
	 L^{\text{actor}}=\mathbb{E}_{\pi_{\theta}}\Big\{\min\left[\Upsilon_\theta A_n, \text{clip}\left(\Upsilon_\theta, 1-\epsilon,1+\epsilon\right)A_n
	\right] \Big\},\label{eq:L_actor}
\end{equation}
where $\mathbb{E}\{\cdot\}$ is the expected value, \text{clip($\cdot$)} is the clip function, $\epsilon$ is a hyperparameter for controlling the range of $\Upsilon_\theta$. In fact, $\epsilon$ is usually a small value that guarantees the policy to be optimized smoothly.

By considering the mean squared error function on the value estimation,
the loss function of the critic network is expressed as
\begin{equation}
	L^{\text{critic}}(\xi)=\left[V^{\xi}(s_{n+1})-V(s_n)\right]^2,\label{eq:L_critic}
\end{equation}
where $V^{\xi}(\cdot)$ is the state value function estimated by critic network and $\xi$ denotes the value parameter. As a result, these networks can be updated according to the gradient of \eqref{eq:L_actor} and \eqref{eq:L_critic}, and old actor is updated by actor for an interval.

\begin{algorithm}[t]
	\caption{PPO-based DRL Training Algorithm}
	\label{alg:PPO}
	\begin{algorithmic}[1]
		\STATE{Initialize network parameters of actor $\theta$, network parameters of critic $\omega$, and replay buffer.}
		\STATE {Initialize  ${\rm ep}=1$.}
		\FOR{${\rm ep}=1 \dots {\rm epl}$}
		\FOR {$n=1\ldots N$}
		\STATE {UAV observes $s_n$ from the environment.}
		\STATE {UAV obtains the action $a_n$ via the actor network.}
		\STATE {Vehicles offload and compute the tasks.}
		\ENDFOR 
		\STATE {UAV synchronizes the transitions $\{s_n,a_n,r_n,s_{n+1}\}$ into DT layer.} 
		\STATE {DT layer calculates reward $r_n,\forall n$.}
		\STATE {Update actor network $\theta$ according to objective function \eqref{eq:L_actor}.}
		\STATE {Update critic network $\omega$ according to loss function \eqref{eq:L_critic}.}
		\STATE {Store the policy entropy and log-probability in the replay buffer.}
		\STATE {DT layer synchronizes actor network to UAV.}
		\ENDFOR
	\end{algorithmic}
\end{algorithm}

The operating environment of PPO algorithm consists of DT model of the whole network environment. UAV observes the state from the DT model and inputs the observed state into the local actor network of PPO algorithm to solve the optimization problem \eqref{c:opt}. Then, the output computation offloading decisions are tested in the DT model and will also feed back to the physical vehicles. The environment information and the actions are periodically synchronized to the DT layer at RSU for reward evaluation, model training, and state monitoring. The training process pseudocode of the proposed PPO framework is given in Algorithm \ref{alg:PPO}.

{\subsection{Complexity Analysis}
The actor and critic networks are composed by multi-layer perceptions (MLPs). For an MLP, the computational complexity of the $j$-th layer is $\mathcal{O}(Z_{j-1}Z_{j}+Z_{j}Z_{j+1})$, where $Z_j$ is the number of neurons for $j$-th layer. Hence, the computational complexity of a $J$-layer MLP is calculated by $\mathcal{O}\left( \sum_{j=2}^{J-1} Z_{j-1}Z_{j}+Z_{j}Z_{j+1} \right)$. 
	Denoting the maximum training episodes and the length of each episode as ${\rm e}^{\max}$ and ${\rm epl}$,  the overall computational complexity for training is calculated by the sum of complexity imposed by actor and critic networks $\mathcal{O}\left({\rm e}^{\max}({\rm epl} \sum_{j=2}^{J-1} Z_{j-1}Z_{j}+Z_{j}Z_{j+1}) \right)$, and for one-step execution is just  $\mathcal{O}\left(  \sum_{j=2}^{J-1} Z_{j-1}Z_{j}+Z_{j}Z_{j+1} \right)$.
}

\begin{figure}[t]
	\centering
	\includegraphics[width=0.4\textwidth]{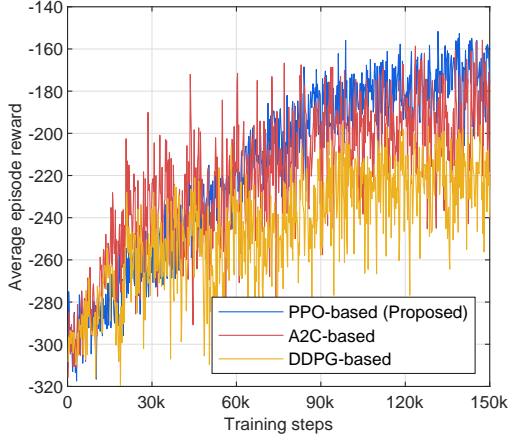}
	\caption{The convergence performance.}
	\label{fig:converge}
\end{figure}

\section{Numerical Results}
In this section, we evaluate the performance of the proposed PPO algorithm for UAV-aided VEC networks. We consider a rectangular areaa of size 500 m $\times$ 500 m, where the vehicles are moving on a cross road with an average velocity of 15 m/s. Unless other stated, we set $K=12$ vehicles. The RSU is located at the (-50 m, 0 m). The UAV is flying at the altitude of $H=100$ m. The channel bandwidth is $B=2$ MHz, the noise power density is $N_0=-130$ dBm/Hz, and the transmit power of vehicles and UAVs are $p_k=0.5$ W and $p_u=0.8$ W, respectively. 
The channel gain is set as $\beta_0=-30$ dB. For computational settings, we have $\kappa=10^{-26}$, $f_{\max}^u=20$ GHz, $D_k[n]\in[0.2 \times 10^6, 2 \times 10^6]$ bits, $C_k[n]\in[500,1500]$ cycles/bit, $T=40$ s, and $N=40$.
The UAV settings $P_0,P_i,U_{\text{tip}},v_0,A$ are set as 39.03 W, 89.07 W, 100 m/s, 3.6 m/s, and 0.5030 $\text{m}^2$, respectively. The maximum acceleration and speed of UAV is $a_{\max}=5$ $\text{m/s}^2$ and $v_{\max}$=20 m/s, respectively. For training settings, the discount factor is $\gamma=0.95$, the length of an episode is equal to $N$, and the penalty factor is $\mu=100$.

\begin{figure}[t]
	\centering
	\includegraphics[width=0.4\textwidth]{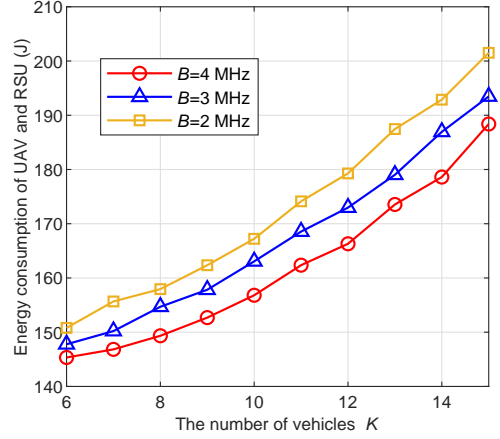}
	\caption{Impact of the number of vehicles and bandwidth.}
	\label{fig:K}
\end{figure}

Fig.~\ref{fig:converge} shows the convergence of reward behavior during the DRL training. We compare the proposed PPO-based method with the deterministic deep policy gradient (DDPG) and the advantage actor-critic (A2C). DDPG is an off-policy DRL algorithm with double actors and double critics, and simply adds exploration noise to output deterministic actions. In contrast, A2C is an efficient on-policy method that substitutes the original reward function with advantage function to better evaluate the quality of state. It can be seen that the proposed PPO approach can efficiently enhance the reward and outperform the DDPG-based method. The proposed PPO algorithm converges at around 100K steps, while DDPG algorithm is more tortuous and reaches the lowest reward and with higher penalty for latency. This verifies that the PPO approach is more steady than the A2C, and can effectively search better policy for the formulated problem than the DDPG.

\begin{figure}[t]
	\centering
	\includegraphics[width=0.4\textwidth]{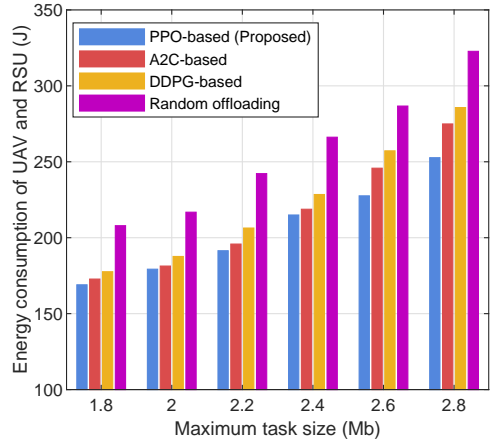}
	\caption{Impact of task size of the vehicles.}
	\label{fig:taskSize}
\end{figure}
\begin{figure}[t]
	\centering
	\includegraphics[width=0.4\textwidth]{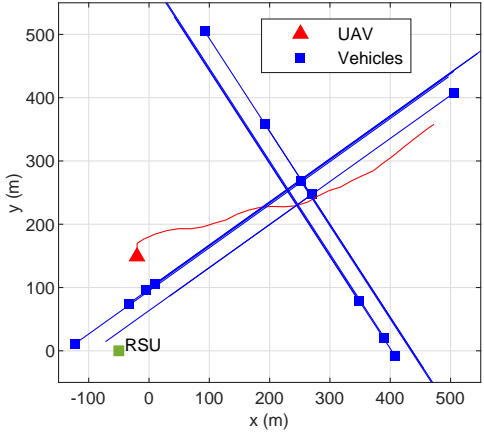}
	\caption{The trajectory of UAV and vehicles.}
	\label{fig:traj}
\end{figure}

To evaluate the impact of number of vehicles, Fig.~\ref{fig:K} presents the objective function versus the number of vehicles under different bandwidth. Intuitively, the energy consumption grows as the number of vehicles increases, and decreases with the increasing of bandwidth. Another observation is that the energy consumption increases faster when more vehicles are served simultaneously. This is because the average computation and communication resources gradually reduce as more vehicles join in the area. Then, the transmission latency increases and more computation resource on UAV is needed by vehicles.

For comparison, we consider three benchmarks in the existing literature, i.e., DDPG algorithm, A2C algorithm, and random offloading.
Note that the random offloading scheme is based on fixed computing frequency allocation and fixed circle trajectory with radius of 300 m at center. It can be found from
Fig.~\ref{fig:taskSize} that the proposed PPO algorithm has the lowest objective, and the random offloading has the highest. As the maximum task size increases, the energy consumption gradually grows faster, and the gap between PPO and DDPG algorithms becomes larger. This is due to the fact that as the task size increases, the computing energy at initial stage of exploiting becomes larger. This makes it hostile for DDPG algorithm, which uses exploration noise to search the action space, to learn more optimal policy than PPO algorithm.

Fig.~\ref{fig:traj} shows the trajectories of vehicles and UAV. We can observe that UAV will quickly fly to the center of target area to reduce the distance between itself and vehicles. By adopting the acceleration model, the trajectory is smooth and is applicable to practical use. Moreover, with DRL control of UAV movement, the decisions of DT become adaptive to unpredictable physical environment. The main reason is that the policy can be preliminarily trained and dynamically adjusts itself to provide the timely optimization for the UAV-aided VEC network.

\section{Conclusion}
This paper proposed a DT framework to realize intelligent offloading in UAV-assisted vehicular networks, where UAV acts both as the edge computing node and the relaying node. We aimed to minimize the system energy consumption performance while ensuring the delay requirement. The state-of-the-art DRL algorithm was designed to obtain near-optimal solution. Numerical results were conducted to demonstrate that the proposed PPO algorithm significantly outperforms the existing benchmarks. 


\bibliographystyle{IEEEtran}
\bibliography{refs}


\end{document}